\documentclass[prb,twocolumn,superscriptaddress,showpacs,amsmath,amssymb]{revtex4}

\usepackage{graphicx}
\usepackage{latexsym}
\usepackage{amsmath}
\usepackage{amssymb}
\usepackage{amsfonts}
\usepackage{color}

\newcommand{\ds}{\displaystyle}

\begin{document}

\title{Vortex States Induced by Proximity Effect in Hybrid \\
Ferromagnet-Superconductor Systems}

\author{A.~V.~Samokhvalov}
\affiliation{Institute for Physics of Microstructures, Russian
Academy of Sciences, 603950 Nizhny Novgorod, GSP-105, Russia}
\author{A.~S.~Mel'nikov}
\affiliation{Institute for
Physics of Microstructures, Russian Academy of Sciences, 603950
Nizhny Novgorod, GSP-105, Russia}
\author{A.~I.~Buzdin}
\affiliation{Institut Universitaire de
France and Universite Bordeaux I, France}

\date{\today}
\begin{abstract}
We consider superconductivity nucleation in multiply connected mesoscopic samples
such as thin-walled cylinders or rings placed in electrical contact with a
ferromagnet. The superconducting critical temperature and
order parameter structure  are studied on the basis of linearized Usadel equations.
We suggest a mechanism of switching between the superconducting states with different
vorticities caused by the exchange field and associated with the oscillatory behavior
of the Cooper pair wave function in a ferromagnet.
\end{abstract}

\pacs{%
74.45.+c, 
74.78.Na, 
74.78.-w  
}

\maketitle

\section{Introduction}

The origin of the vortex states in hybrid
ferromagnet (F)-superconductor (S) structures is closely related to the
basic mechanisms responsible for the interplay
between the ferromagnetic and superconducting orderings
(see, e.g., Ref.~\cite{Buzdin-RMP05}).
The first mechanism is associated with the orbital effect, i.e.,
interaction of Cooper pairs with the magnetic field induced
by magnetic moments
\cite{Ginzburg-JETP56}.
This field causes the appearance of inhomogeneous
superconducting phase distributions and spontaneous vortex states.
The switching between these states characterized by different winding numbers
can result in an oscillatory behavior of
 the critical temperature $T_{c}$ as a function of the external field $H$, which
resembles the Little-Parks effect in multiply connected
superconducting samples \cite{Little-Parks-PR64}.
Such nonmonotonic behavior of $T_{c}(H)$
was shown to be inherent to hybrid F/S systems
with magnetic dots or domains \cite{Lange-PRL03,Aladyshkin-JCM03,Aladyshkin-PRB03},
that create a "magnetic template" for nucleation of
superconducting order parameter.
The second mechanism arises from the exchange interaction which comes into play because of
 the proximity effect, when the Cooper pairs penetrate
into the F layer and induce superconductivity there.
The latter effect is known to result in the
 damped-oscillatory behavior of the Cooper pair wave function
in a ferromagnet
\cite{Buzdin-RMP05}
which is the cause of a number of fascinating interference phenomena in hybrid F/S
structures.
In particular, this peculiar proximity effect reveals itself in the
 oscillating \cite{Buzdin-JETPL90,Radovic-PRB91}
or re-entrant \cite{Tagirov-PhC98}
behavior of the critical temperature  as
a function of a ferromagnetic layer thickness in layered F/S structures,
and is responsible for the formation of $\pi$-junctions
\cite{Buzdin-JETPL82,Buzdin-JETPL91,Ryazanov-PRL01}.
A Josephson $\pi$-junction is a generic example of the system where the proximity
effect in a ferromagnetic subsystem is used to obtain an energetically favorable
superconducting state with a nontrivial distribution of the order parameter phase.
A resulting distinctive feature of the systems with $\pi$-junctions is a possible
 unusual ground state with spontaneous supercurrents and, in particular, with
 spontaneously formed vortices.
For a  Josephson  junction with a step-like change in the F layer thickness
such spontaneously formed vortex states have been discussed,
 e.g., in Refs.~\cite{Bulaevsky-SSC78,Goldobin-PRL06,Frolov-PRB06}.

It is the purpose of this paper to examine a possibility to realize a switching
between the spontaneously created vortex states in multiply connected samples caused by the proximity effect with
a ferromagnet. We focus on the behavior of critical temperatures for superconducting states with different
vorticities and, thus, in some sense study an analog of the Little-Parks effect
caused by the exchange interaction mechanism.

\begin{figure}[b!]
\includegraphics[width=0.17\textwidth]{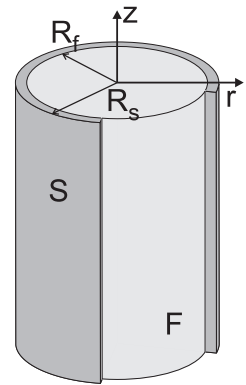}
\caption{The schematic representation of the F/S system
under consideration:
thin-walled superconducting shell around a ferromagnetic cylinder.
Here $R_f$ is the radius of the F core, and $R_s$ is the
outer radius of the S shell, $(r,\theta,z)$ is the cylindrical
coordinate system.}
\label{Fig:1}
\end{figure}
%

\begin{figure}[t!]
\includegraphics[width=0.45\textwidth]{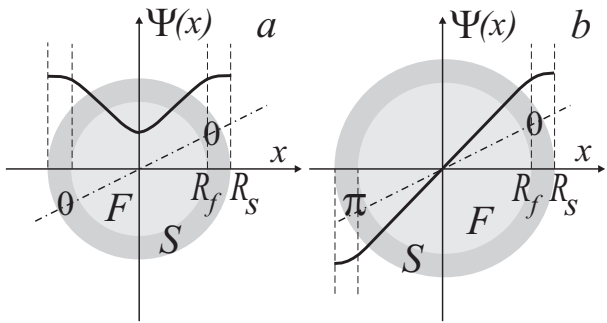}
\caption{The schematic behavior of the superconducting
order parameter inside the F cylinder.
(a) The curve $\Psi(x)$ represents sketchy
the behavior of the pair wave function in the $L=0$
phase. Due to symmetry the derivative $\partial_x \Psi$
is zero at the center of F cylinder.
(b) The pair wave function in the
phase with $L = 1$ vanishes at the center of F cylinder,
and $\Psi(x)$ has a $\pi-$shift in diametrically
opposite points.}
\label{Fig:2}
\end{figure}

To elucidate our main results we start from a qualitative discussion
of the proximity effect on the superconducting ground state in
a thin-walled superconducting
shell surrounding a cylinder (core) of a ferromagnetic metal
(see Fig.~\ref{Fig:1}).
We expect that the superconducting ground state
in such geometry should be strongly influenced by the
 damped-oscillatory behavior of the superconducting
order parameter in ferromagnetic cylinder
and, thus, controlled by the ratio of
the period of the order parameter
oscillations ($\sim \xi_f$) to the radius $R_f$ of the F core.
Indeed, for $R_f < \xi_f$, the variation of the pair wave function $\Psi$
along a line crossing
the F core appears to be modest and
the order parameter can not  change its sign along the line.
It means that in the ground state the superconducting phase
in diametrically opposite points must be the same.
The resulting angular momentum $L$ of the
pair wave function $\Psi$ is equal to zero (see Fig.~\ref{Fig:2}a).
This state is analogous to the $0-$phase state of
SFS layered structures
\cite{Buzdin-JETPL91}.
%
%
For a larger radius $R_f \gtrsim \xi_f$,
the pair wave function may cross zero at the axis
of the F cylinder  which causes a $\pi-$shift
in the superconducting phase
in diametrically opposite points.
In this case the angular momentum $L$ of the pair
wave function $\Psi$ is nonzero (Fig.~\ref{Fig:2}b).
Thus, the penetration of Cooper pairs into the F core
and exchange interaction can induce the superconducting states
with a nonzero vorticity.
It is natural to expect that the supercurrents flowing in
 such states  with $L \neq 0$ are responsible for an additional
contribution to the vortex energy.
The interplay between the exchange effect and supercurrent depending
energy term
 may result in a subsequent switching between the
states with different vorticities, as the F core radius
increases. An obvious consequence of
these transitions between the states with different $L$
should be a nonmonotonic dependence of
the critical temperature $T_c$
on the F core radius and exchange field.

The paper is organized as follows.
In Sec.~\ref{ModelSection} we briefly discuss the basic equations.
In Sec.~\ref{ResultsSection} we study the switching between
different vortex states in two model F/S systems.
The first system consists of a thin-walled superconducting
cylindrical shell surrounding a cylinder of
a ferromagnetic metal.
The second one is a planar structure which consists of
a mesoscopic superconducting ring placed
at the surface of a thin ferromagnetic film.
For both cases we assume that there is a good electrical
contact between the F and S regions, to assure a rather strong proximity
effect.
We summarize our results in Sec.~\ref{DiscusSection}.


\section{Model}\label{ModelSection}

We assume the elastic electron-scattering time $\tau$
to be rather small, so that the critical temperature $T_c$ and
exchange field $h$ satisfy the dirty-limit conditions
$T_c\tau \ll 1$ and $h\tau \ll 1$.
In this case a most natural approach to calculate $T_c$
 is based on the Usadel equations
\cite{Usadel-PRL70}
for the averaged anomalous Green's function $F_{f}$ and $F_{s}$
for the F and S regions, respectively (see \cite{Buzdin-RMP05}
for details).
Near the second-order  superconducting phase transition,
 the Usadel equations can be linearized with respect to
the pair potential $\Delta(\mathbf{r})$.
In the F (S) region these linearized Usadel equations
take the form
\begin{eqnarray}
    & &-\frac{D_f}{2} \nabla^2 F_f
    + (\vert\, \omega\, \vert + \imath\, h\, {\rm sgn}\, \omega) F_f
    = 0\,,                                      \label{eq:1} \\
    & &-\frac{D_s}{2} \nabla^2 F_s + \vert\, \omega\, \vert F_s
    = \Delta(\mathbf{r})\,.                     \label{eq:2}
\end{eqnarray}
The superconducting critical temperature $T_c$
is determined from the self-consistency condition for the gap function:
\begin{equation} \label{eq:5}
    \Delta(\mathbf{r})\,\ln \frac{T_c}{T_{c0}}
    + \pi T_c \sum_\omega \left(\frac{\Delta(\mathbf{r})}{\vert \omega \vert}
        - F_s(\mathbf{r},\omega)\right) = 0.
\end{equation}
Here $D_{f}$  and $D_{s}$ are the diffusion constants in the ferromagnet
and superconductor, respectively, and
$\omega=(2n+1)\pi T_c$ is a Matsubara frequency at the temperature $T_c$.
Equations (\ref{eq:1}),(\ref{eq:2}) must be supplemented with
the boundary condition at the outer surfaces
\begin{equation} \label{eq:3}
     \partial_{\mathbf{n}} F_{f,s} = 0\,,
\end{equation}
and at the interface between the F and S metals:
\cite{Kuprianov-JETP88}
\begin{equation} \label{eq:4}
    \sigma_s\, \partial_{\mathbf{n}} F_s
    = \sigma_f\, \partial_{\mathbf{n}} F_f; \quad
    F_s = F_f + \gamma_b \xi_s\, \partial_{\mathbf{n}} F_f\,.
\end{equation}
Here $\xi_s=\sqrt{D_s/2\pi T_{c0}}$ is the superconducting coherence length,
$\sigma_{f}$ and $\sigma_{s}$ are the normal-state conductivities of the F and S metals,
$\gamma_b$ is related to the F/S boundary resistance $R_b$ per unit area
through $\gamma_b \xi_s = R_b \sigma_f$, and $\partial_{\mathbf{n}}$
denotes a derivative taken in the direction perpendicular to the outer surfaces
or to the F/S interface.
For the sake of simplicity we assume $h \gg \pi T_{c0}$
and neglect the proximity effect suppression caused by
a finite F/S interface resistance %
\cite{Fominov-PRB02}, %
i.e., take $\gamma_b \to 0$.
In this regime we get $F_f = F_s$ at the F/S interface.

For a system with a cylindrical symmetry the vorticity parameter $L$
just coincides with the angular momentum of
the Cooper pair wave function. Choosing cylindrical coordinates ($r,\theta,z$)
we look for solutions of the equations
(\ref{eq:1}),(\ref{eq:2}),(\ref{eq:5}) characterized by certain
angular momenta $L$:
\begin{equation} \label{eq:6}
   \Delta(\mathbf{r})=\Delta(r,z)\, {\rm e}^{\imath L \theta}, \quad
   F_{f,s}(\mathbf{r})=f_{f,s}(r,z)\, {\rm e}^{\imath L \theta},
\end{equation}
According to the equations (\ref{eq:1}),(\ref{eq:2}),(\ref{eq:5})
there is a symmetry $F_{f,s}(\omega)=F_{f,s}^*(-\omega)$, so that
we can treat only positive $\omega$ values.
The Usadel equations (\ref{eq:1}),(\ref{eq:2}) can be written
in the form
\begin{eqnarray}
    & &-\frac{D_f}{2} \left(\frac{1}{r}\,\partial_r ( r\,\partial_r f_f )
       + \partial_z^2 f_f -\frac{L^2}{r^2} f_f\,\right)   \label{eq:7} \\
    & &\qquad\qquad\qquad  +\imath\, h\, f_f = 0, \nonumber \\%
    & &-\frac{D_s}{2} \left(\frac{1}{r}\,\partial_r ( r\,\partial_r f_s )
       + \partial_z^2 f_s - \frac{L^2}{r^2} f_s\, \right) \label{eq:8} \\
    & &\qquad\qquad\qquad  + \omega\, f_s = \Delta\,. \nonumber %
\end{eqnarray}
An appropriate self-consistency equation (\ref{eq:5})
can be rewritten as follows:
\begin{equation} \label{eq:9}
    \Delta\,\ln\frac{T_c}{T_{c0}}
    + 2 \pi T_c \sum_{\omega>0} \left(\frac{\Delta}{\omega}
        - {\rm Re}\, f_s(\omega)\right) = 0.
\end{equation}
%


\section{Critical Temperature of Vortex States}
\label{ResultsSection}

Now we proceed with the critical temperature calculations for different vortex states.
For the sake of definiteness we consider here two generic examples of hybrid F/S systems
which we believe to manifest the vorticity switching scenario suggested in the introduction.

\subsection{Thin-walled superconducting shell
around a ferromagnetic cylinder}\label{ShellSection}
\label{ResultsSection-a}

Consider a superconducting cylindrical shell of a
thickness $W = R_s - R_f\ll R_f$ surrounding a thin cylinder of
a ferromagnetic metal  with a uniform magnetization
$\mathbf{M} = M \mathbf{z}_0$.
Here $R_f$ is the radius of the F core, and $R_s$ is the
outer radius of the S shell (see Fig.~\ref{Fig:1}).
Naturally, to observe the pronounced influence of the
proximity effect on the transition temperature,
the thickness of the S shell $W$ must be smaller
than the superconducting coherence
length $\xi_s$.

For a thin and long F cylinder with
the magnetization direction chosen along the $z-$axis
the magnetic field $B$ vanishes outside the $F$ region.
However, this magnetization induces
a vector potential $A_\theta \simeq 2\pi M R_f$
in the S region, which may result in a standard Little-Parks effect of
 electromagnetic origin.
We may take account of this vector potential in Eq.~(\ref{eq:8})
 replacing
the vorticity parameter $L$ by the value $L - \Phi / \Phi_0$,
where $\Phi_0=2\pi\hbar c / e$ is the magnetic flux
 quantum and $\Phi = 4\pi^2 M R_f^2$ is the total magnetic flux.
For the standard Little-Parks effect the critical temperature vs $\Phi$
 oscillates with a period $\Phi_0$ and an amplitude $\Delta T_c\sim T_c\,\xi_s^2/R_f^2$.
 For typical parameters $M \sim {\rm 10^2\,G}$, $T \sim {\rm 10\,K}$,
$D_s \sim {\rm 10\, cm^2/s}$ and $R_f$ of order of several
 $\xi_f \sim 10\,nm$ lengths we get $\Phi / \Phi_0\ll 1$ and
 $\Delta T_c\ll T_c$. These simple estimates allow us to find a region of parameters
 where we can exclude the effect of magnetic field on $T_c$.
The above effect of the magnetic field can be also weakened provided we decrease the
height of the F/S cylinder going over to the case of a thin disk when the field $B$
is suppressed due to the demagnetization factor.
 Note that choosing the magnetization direction in the plane perpendicular to the
 cylinder axis we can get rid of the standard Little-Parks effect completely because
 of the absence of the magnetic field component along the cylinder axis.
On the contrary, the vorticity switching scenario studied below does not depend on the
magnetization direction.

We look for homogeneous along $z$ solutions of the equations
(\ref{eq:7}), (\ref{eq:8}) with a certain
angular momentum $L$.
In this case  the equation (\ref{eq:7}) in the F cylinder
can be readily solved:
\begin{equation}\label{eq:11}
    f_f = C\, I_L(q_f r), \quad q_f = \frac{1 + \imath}{\xi_f}\,.
\end{equation}
Here $I_L(u)$ is the modified Bessel function of first kind of order $L$,
and $\xi_f=\sqrt{D_f / h}$ is the characteristic length scale of the order parameter variation
 in the F metal.
In the dirty limit parameter $\xi_f$ determines both the length scale
of oscillations and decay length for the Cooper pair wave function
in a ferromagnet \cite{Buzdin-RMP05}. The boundary conditions (\ref{eq:3}),(\ref{eq:4})
for Eq.~(\ref{eq:8}) take the form:
\begin{eqnarray}\label{eq:12}
   & &\sigma_s\,\frac{d f_s}{dr}\,{\bigg\vert_{R_f}} %
      = \alpha_L\,q_f\sigma_f f_s(R_f)\,, \quad
      \frac{d f_s}{dr}\,{\bigg\vert_{R_s}}=0, \\
   & &\quad\alpha_L =\frac{L}{u_f}+ \frac{I_{L+1}(u_f)}{I_L(u_f)}\,,\:\:\:
        u_f = q_f R_f. \nonumber
\end{eqnarray}
%
%
\begin{figure}[t!]
\includegraphics[width=0.4\textwidth]{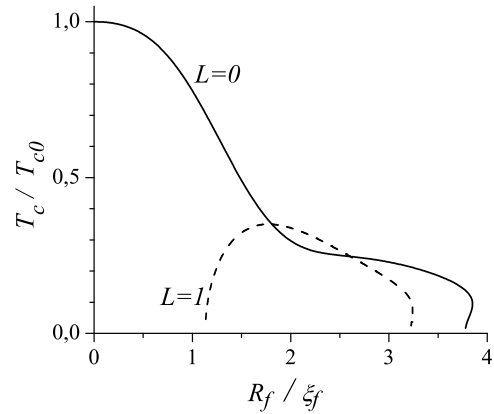}
\caption{The dependence of the critical temperature $T_c$
on the  F core radius $R_f$ for two values of the vorticity
$L = {\rm 0}$ (solid line) and  $L = {\rm 1}$ (dashed line).
Here we choose
$W = {\rm 0.5} \xi_s$; $\sigma_s / \sigma_f = {\rm 2.5}$;
$\xi_s / \xi_f = {\rm 0.265}$.}
\label{Fig:3}
\end{figure}
%
\begin{figure*}
\includegraphics[width=0.4\textwidth]{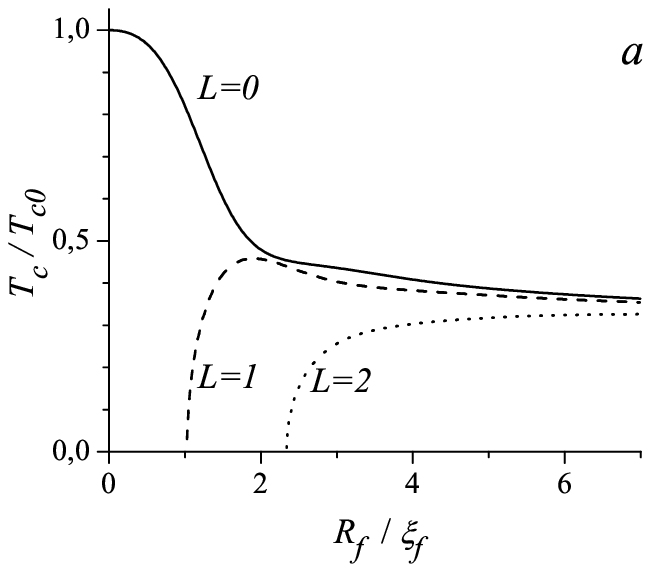}
\includegraphics[width=0.4\textwidth]{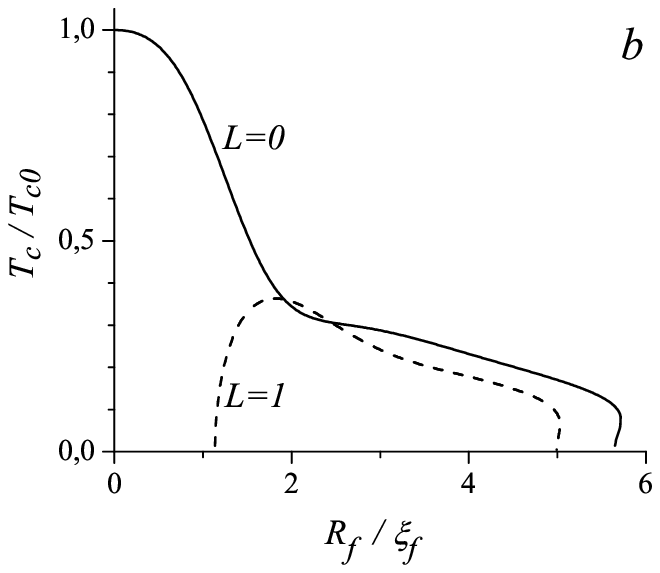}
\includegraphics[width=0.4\textwidth]{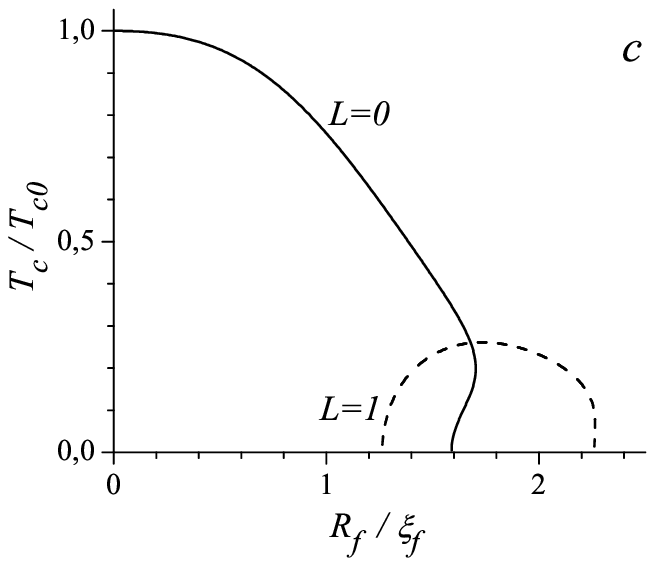}
\includegraphics[width=0.4\textwidth]{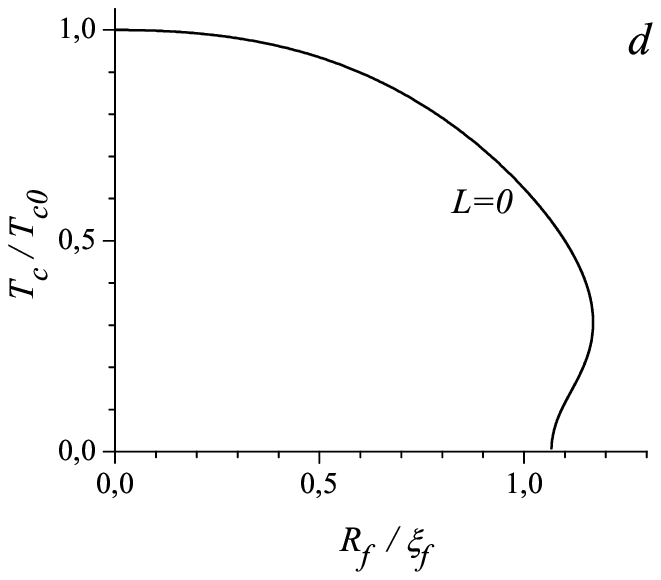}
\caption{The typical dependences of the critical
temperature $T_c$ on the F core radius $R_f$
for different values of the vorticity $L = {\rm 0}$ (solid line),
$L = {\rm 1}$ (dashed line) and , $L = {\rm 2}$ (dotted line).
Here we choose $W= {\rm 0.5} \xi_s$; $\xi_s / \xi_f = {\rm 0.28}$
and different values of the ratio $\sigma_s / \sigma_f$:
a) $\sigma_s / \sigma_f = {\rm 3}$;
b) $\sigma_s / \sigma_f = {\rm 2.7}$;
c) $\sigma_s / \sigma_f = {\rm 2.5}$;
d) $\sigma_s / \sigma_f = {\rm 2}$.}
\label{Fig:4}
\end{figure*}
For $W \ll \xi_s$, the variations of the functions $f_s(r)$ and $\Delta(r)$ in
the superconducting shell are small: $f_s(r)\simeq f$, $\Delta(r)\simeq \Delta$.
Therefore, we can average Eq.~(\ref{eq:8})
over the thickness of the S shell, using the
boundary conditions (\ref{eq:12}) to integrate the term
$\partial_r(r\,\partial_r f_s)$.
Finally, we obtain the following expression:
\begin{equation}\label{eq:13}
    f = \frac{\ds \Delta}{\ds \omega + \frac{D_s}{2 } %
       \left[\left(\frac{L}{R_f}\right)^2
       + \frac{\sigma_f\,q_f}{\sigma_s\, W} \alpha_L \right] }.
\end{equation}
Substituting  Eq.(\ref{eq:13}) into Eq.(\ref{eq:9})
one obtains a self-consistency equation for the critical
temperature $T_c$:
\begin{equation}\label{eq:14}
    \ln\frac{T_c}{T_{c0}} = %
        \Psi\left(\frac{1}{2}\right)
          - Re\,\Psi\left(\frac{1}{2}+\Omega_L\right),
\end{equation}
where $\Psi$ is the digamma function.
The depairing parameter
\begin{equation}\label{eq:15}
    \Omega_L = \frac{1}{2}\frac{T_{c0}}{T_c}\,\xi_s^2 %
                    \left[\left(\frac{L}{R_f}\right)^2 %
                    + \frac{\sigma_f\,q_f}{\sigma_s W}\, %
                                \alpha_L \right]
\end{equation}
is responsible for the superconductivity destruction in the shell
due to both the exchange effect and the supercurrent flowing around the cylinder.

Figure~\ref{Fig:3} shows a typical dependency of the critical
temperature $T_c$
on the F core radius $R_f$,
obtained from Eqs.~(\ref{eq:14}),(\ref{eq:15})
for  different winding numbers $L$.
We see that for a small F cylinder radius $R_f \ll \xi_f$
only the state with $L=0$ appears to be energetically favorable.
The influence of the proximity effect is weak
and the critical temperature $T_c$ is close to $T_{c0}$.
The $T_c$ of a vortex state with $L \ne 0$ is suppressed
because of a large supercurrent energy.
The increase in the radius $R_f$ results in
a decrease in $T_c$ for the state with $L=0$ and reduce
the kinetic energy of supercurrents for $L \ne 0$.
At the same time, the damped-oscillatory behavior
of the superconducting order parameter in a ferromagnet
becomes important.
If the diameter of the F cylinder is comparable with
the period of the order parameter oscillations ($\sim \xi_f$)
and satisfies approximately  the conditions of the
$\pi-$phase superconductivity in layered F/S structures
($2\xi_f < 2R_f < 5\xi_f$)
\cite{Buzdin-RMP05},
then there appears a $\pi-$shift in the phase of the
superconducting order parameter
in diametrically opposite points (see Fig.~\ref{Fig:2}b).
In this case, the critical temperature of $L=1$ state
becomes higher than the critical temperature
of the state with $L=0$.
Thus, our calculations confirm the qualitative arguments given in the introduction.
The penetration of Cooper pairs into the F core
and the phase shift of the pair wave function $\Psi$
due to the exchange interaction can induce vortex states
in the superconducting shell.

To illustrate the scenario of switching between the states with different
vorticities $L$ and a nonmonotonic dependence of the critical temperature $T_c$
vs $R_f$ we present here several $T_c(R_f)$ curves for various conductivity ratio
(see Fig.~\ref{Fig:4}).
We see that the ratio $\sigma_{s}/\sigma_{f}$ of the normal-state conductivities
 of the F and S metals is an important factor,
controlling the generation of the vortex states
in the F/S structure under consideration.


\subsection{Superconducting ring  at the surface of a
thin ferromagnetic film}
\label{ResultsSection-b}

\begin{figure}
\includegraphics[width=0.4\textwidth]{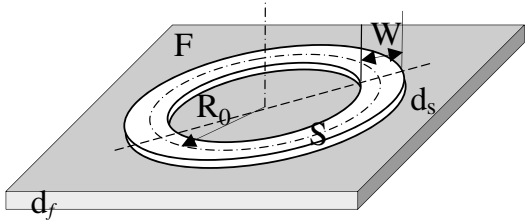}
\caption{Superconducting ring lying at the surface of a
thin ferromagnetic film.
Here $R_0$ and $W$ are the radius and the width
of the S ring, and $d_{f}$ ($d_{s}$) is the
thickness of a ferromagnetic (superconducting) layer.}
\label{Fig:5}
\end{figure}

As a second example we consider a planar hybrid system, i.e.,
a superconducting ring lying at a ferromagnetic film
with a uniform in-plane magnetization $\mathbf{M}$
(see Fig.~\ref{Fig:5}).
Such version of the setup can be more convenient
for the experimental observation of the switching phenomena.
%
%
%
\begin{figure*}[t!]
\includegraphics[width=0.4\textwidth]{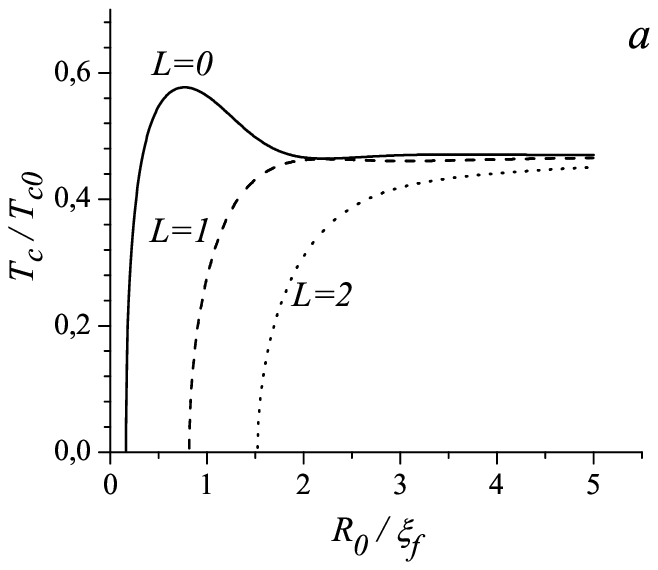}
\includegraphics[width=0.4\textwidth]{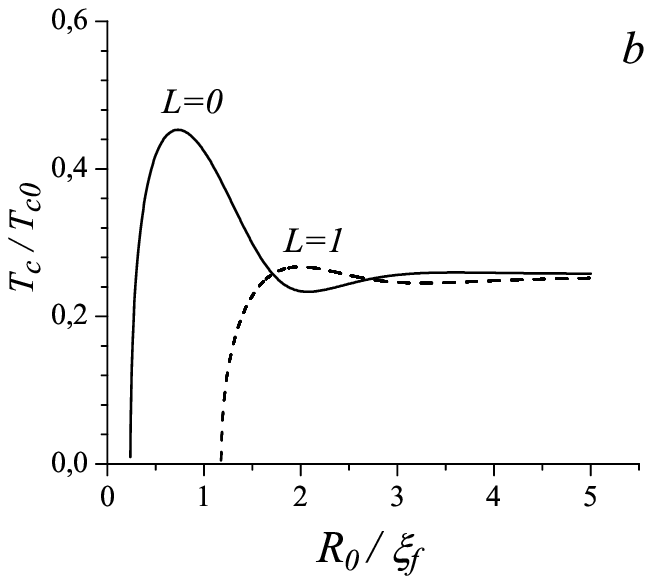}
\includegraphics[width=0.4\textwidth]{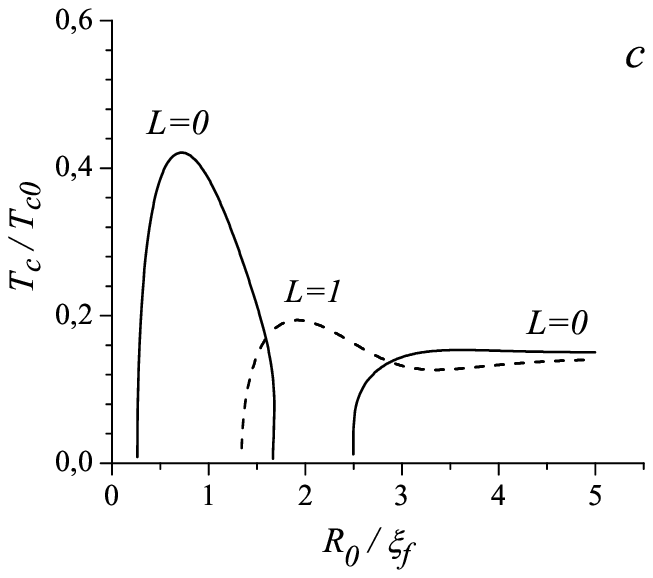}
\includegraphics[width=0.4\textwidth]{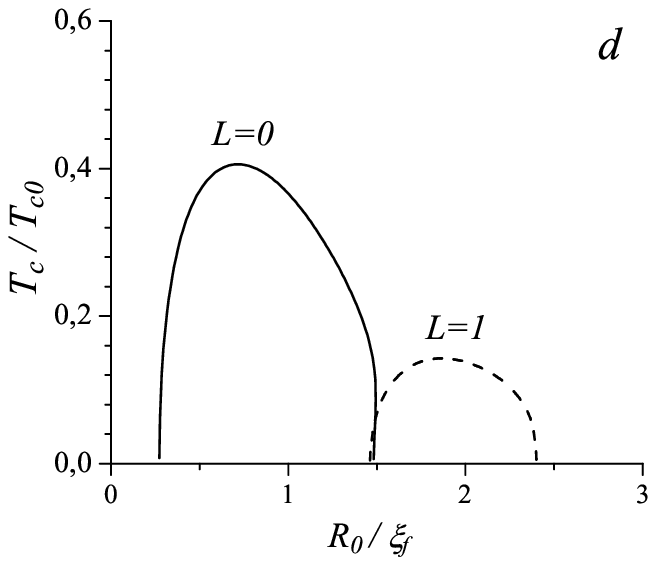}
\caption{The dependence of the critical temperature $T_c$
on the S ring radius $R_0$ for different
values of the vorticity $L = {\rm 0}$ (solid line),
$L = {\rm 1}$ (dashed line) and , $L = {\rm 2}$ (dotted line).
Here we choose $d_s / d_f = {\rm 1}$, $W = {\rm 0.5} \xi_s$,
$\xi_s / \xi_f = {\rm 0.1}$:
a) $\sigma_s / \sigma_f = {\rm 2.5}$; %
b) $\sigma_s / \sigma_f = {\rm 2.1}$;
c) $\sigma_s / \sigma_f = {\rm 2.03}$;
d) $\sigma_s / \sigma_f = {\rm 2.0}$.}
\label{Fig:6}
\end{figure*}
%
The superconducting ring of the radius $R_0$ and the width $W \ll R_0$
occupies the region
$R_0 - W/2 < r < R_0 + W/2$, $0 < z < d_s$.
The boundary conditions at the interfaces with vacuum yield
$\partial_r f_s(R_0 \pm W/2,z)=0$, and $\partial_z f_s(r,d_s)=0$.
For simplicity, we consider only the case of a rather thin ring with
$W \ll \xi_{f,s}$ and $d_s \ll \xi_s$ which allows us to
assume  the variations of the functions $f_s(r)$ and $\Delta(r)$ in
the superconducting ring to be small.
Thus,
we can average Eq.~(\ref{eq:8}) over the volume of the S ring,
integrate the terms $\partial_r(r\, \partial_r f_s)$ and
$\partial^2_z f_s$, and make use of the boundary condition
at the interface with vacuum.
Finally, we get the following expression for the derivative
$\partial_z f_s$ at $r=R_0,\, z=0$:
\begin{equation}\label{eq:16}
    \frac{1}{d_s}\,\partial_z f_s = %
        \frac{2}{D_s}\left(\Delta - \omega\,f_s\right) %
        - \left(\frac{L}{R_0}\right)^2\! f_s\,.
\end{equation}

The F metal occupies the region $-d_f < z < 0$.
We will address the case of a very thin F film: $d_f \ll \xi_f$.
The boundary conditions at the interfaces with vacuum yield
$\partial_z f_f=0$.
Far from the superconducting ring, i.e. for $r-R_0-W/2\gg d_f$ and
for $R_0-W/2-r\gg d_f$ we can average
the Usadel equation  (\ref{eq:7})
 over the thickness $d_f$:
\begin{equation}
   -\frac{D_f}{2}\,\left(\frac{1}{r}\,\partial_r ( r\,\partial_r f_f )
       -\frac{L^2}{r^2} f_f\,\right)
       + \imath\,h f_f = 0 \ . \label{eq:17}
    \end{equation}
The solution of this equation reads:
\begin{equation}\label{eq:19}
    f_f(r) = \left\{
            \begin{array}{ccc}
                C_1\, I_L (q_f r)& , & R_0-W/2-r\gg d_f \\
                \\
                C_2\, K_L (q_f r)& , & r-R_0-W/2\gg d_f
            \end{array} \right.
\end{equation}
where $I_L(u)$ and $K_L(u)$ are the modified Bessel functions
of order $L$.
Making use of this solution one can easily get
the following relations between
the function $f_f$ and  its derivative
$\partial_r f_f$ at $r=R_0\pm \varepsilon$:
\begin{eqnarray}
    & &\frac{d f_f}{dr}\bigg\vert_{R_{0}-\varepsilon} =
       q_f \left( \frac{L}{u_0} + \frac{I_{L+1}(u_{0})}{I_{L}(u_{0})}%
                       \right)f_f(R_{0}-\varepsilon)\,,\qquad \label{eq:20}\\
    & &\frac{d f_f}{dr}\bigg\vert_{R_{0}+\varepsilon} =
       q_f \left( \frac{L}{u_0} - \frac{K_{L+1}(u_{0})}{K_{L}(u_{0})}%
                       \right)f_f(R_{0}+\varepsilon)\,,\qquad \label{eq:21}
\end{eqnarray}
where $u_{0} = q_f R_{0}$ and $\mbox{max}[W,d_f]\ll\varepsilon\ll\xi_f\lesssim R_0$.
Assuming $f_f$ to be a slow function of $r$ one can write a boundary condition
on the derivative jump:
\begin{eqnarray}
    & &\frac{d f_f}{dr}\bigg\vert_{R_{0}-\varepsilon}^{R_{0}+\varepsilon} \simeq
        - q_f\, Q_L\, f_f(R_0)\,,                 \label{eq:22} \\
    & &\quad Q_L = \frac{I_{L+1}(u_0)}{I_{L}(u_0)} +
          \frac{K_{L+1}(u_0)}{K_{L}(u_0)}\,.    \nonumber
\end{eqnarray}
On the other hand in the region $|r-R_0|\ll \xi_f$ the Eq.~(\ref{eq:7})
takes a simple form:
$$
\frac{1}{r}\,\partial_r ( r\,\partial_r f_f )
       + \partial_z^2 f_f =0 \ .
$$
 Integrating this equation
over the region $R_0-\varepsilon<r<R_0+\varepsilon$ and over the ferromagnetic
film thickness and making use of the boundary conditions described above we obtain:
\begin{equation}\label{eq:23}
   W\,\partial_z f_f\bigg\vert_{z=0,r=R_0} = q_fd_f\, Q_L  f_f (R_0)
\end{equation}
As before we restrict ourselves to the case of low F/S interface
resistance assuming $\gamma_b = 0$ in (\ref{eq:4}).
In this regime, $f_f(R_0,0) \simeq f_s(R_0,0) \equiv f$.
The Eqs.~(\ref{eq:16}),(\ref{eq:23}) and the boundary conditions
at the F/S interface (\ref{eq:4}) determine the amplitude $f$:
\begin{equation}\label{eq:24}
    f = \frac{\Delta}{\ds \omega + \frac{D_s}{2}%
            \left[  \left(\frac{L}{R_0}\right)^2
              +\frac{ q_f}{\eta W}\, Q_L \right]}\,,
\end{equation}
where $\eta=\sigma_s d_s / \sigma_f d_f$.
Substitution of (\ref{eq:24}) into Eq.(\ref{eq:9})
results in the self-consistency equation for the critical
temperature $T_c$ of the F/S hybrid (\ref{eq:14}),
where the depairing parameter $\Omega_L$ is determined by the
following expression:
\begin{equation}
    \Omega_L = \frac{1}{2}\frac{T_{c0}}{T_c}\,%
               \xi_s^2 %
                \left[\,
                      \left(\frac{L}{R_0}\right)^2%
                \right.
                \left. +\frac{q_f}{W\eta}\, Q_L(u_0)\,%
                \right].  \label{eq:25}
\end{equation}

In Fig.~\ref{Fig:6}, we present typical
dependences of the critical temperature $T_c$
on the S ring radius $R_0$ for the different orbital number $L$,
obtained from Eqs. (\ref{eq:14}), (\ref{eq:25}).
The curves appear to be qualitatively similar to the ones obtained for a
superconducting thin-walled cylinder in the previous subsection.
Note that for a particular choice of parameters (see Fig.\ref{Fig:6}c)
 $T_c$ vanishes in a certain interval of $R_0$ values and
we observe an interesting re-entrant behavior of the critical temperature
for a state with zero vorticity.

\section{Summary}
\label{DiscusSection}
To summarize, we suggest a mechanism of switching between the superconducting states with different
vorticities caused by the exchange field in the hybrid S/F structures
 and associated with the oscillatory behavior
of the Cooper pair wave function in a ferromagnet.
We defined the range of the system parameters at which the predicted effect can be
experimentally observable. The most
restrictive condition is imposed on the relation between the superconducting cylinder or ring radius
$R_0$ and coherence lengths $\xi_s$ and $\xi_f$. On the one hand the radius should
essentially
exceed the superconducting coherence length to decrease the kinetic energy of
supercurrents in the states with nonzero vorticity,
 but on the other hand it should be of the order of only several $\xi_f$
lengths to ensure a rather strong influence of the proximity effect.
As a result, to observe the switching effect we need to consider the S/F systems with
a rather large ratio $\xi_f/\xi_s$.
This ratio can be increased if we choose a superconducting material with a rather
short coherence length, e.g., heavy fermion compounds \cite{heavy}.
Despite of such restriction we believe that the proximity induced switching between the vortex states
can be experimentally observable and would provide an interesting manifestation of
interference effects in mesoscopic superconductivity.

\section{Acknowledgements}

We are indebted to A.\ A.\ Fraerman for useful discussions.
This work was supported by the bilateral project BIL/05/25
between Flanders and Russia, by the Russian Foundation for Basic
Research, by the program of Russian Academy of Sciences "Quantum
Macrophysics", and by the Russian Science Support foundation (A.S.M.).

\end{document}